\documentclass{main}

\usepackage{textcomp}
\usepackage{fancyhdr}
\usepackage{lastpage}
\usepackage{caption}
\usepackage{subcaption}
\usepackage{amsmath}
\usepackage{cleveref}
\usepackage{siunitx}
\usepackage[absolute]{textpos} 
\def\be{\begin{equation}}
\def\ee{\end{equation}}
\def\bea{\begin{eqnarray}}
\def\eea{\end{eqnarray}}

\newcommand{\powhegbox}{{\tt POWHEG\,BOX}}
\crefname{figure}{Fig.}{Figs.}

\fancypagestyle{firstpagefooter}{%
  \fancyfoot[L]{  \textcopyright \footnotesize This work is an open access article under a Creative Commons Attribution 4.0 International License (https://creativecommons.org/licenses/by/4.0/).}
  \fancyfoot[C]{}  
   
}

\begin{document}

\thispagestyle{firstpagefooter}
\title{\Large Dijet Photoproduction in \powhegbox}

\author{\underline{A.~Feike}\,\footnote{Speaker, email: alex.feike@uni-muenster.de} and T.~Je\v{z}o and M.~Klasen}

\address{
Institut für Theoretische Physik, Universität Münster, Wilhelm-Klemm-Straße 9, 48149 Münster, Germany
}

\begin{textblock*}{50mm}(156mm,5mm)
  MS-TP-25-25
\end{textblock*}


\maketitle
\abstracts{
Photoproduction processes have gained a renewed interest following the approval of the EIC, making their implementation in Monte Carlo event generators highly desirable. 
We present recent efforts to develop a \powhegbox\ extension simulating dijet production from direct and resolved photons at next-to-leading order in QCD merged to parton showers, employing the Weizsäcker-Williams Approximation.
It will facilitate event generation for collisions involving leptons, protons and heavy ions.
Thus, it will be particularly useful for the study of ultra-peripheral collisions at CERN’s LHC and for making predictions relevant to BNL's EIC.
}




\section{Introduction}
\begin{figure}[h]
     \centering
     \begin{subfigure}[b]{0.4\textwidth}
         \centering
         \includegraphics[width=\textwidth]{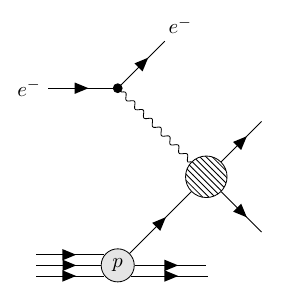}
         \caption{}
         \label{fig:direct}
     \end{subfigure}
     \begin{subfigure}[b]{0.4\textwidth}
         \centering
         \includegraphics[width=\textwidth]{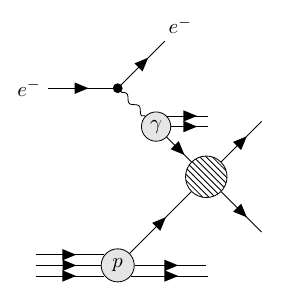}
         \caption{}
         \label{fig:resolved}
     \end{subfigure}
        \caption{Direct (a) and resolved (b) dijet photoproduction in $ep$ collision.}
        \label{fig:diagram}
\end{figure}

Jet production is an important process at high-energy colliders and calculable in perturbative QCD.
For $ep$ collisions, the relevant Feynman diagrams are sketched in \cref{fig:diagram}.
Here, the incoming lepton emits a photon with low virtuality, which then interacts with the partons of the proton beam.
This can happen either directly, with the photon coupling to the hard subprocess, or the photon develops a hadronic structure, giving rise to the resolved photoproduction channel.
\\
In the hard interaction, we consider the production of two jets at leading order, with additional jets originating from real emissions, multi-parton interactions, and spectator remnants from the hadron or the resolved photon.
Since we focus on photons with low virtuality, this process is complementary to deep-inelastic scattering, which involves highly virtual photons.
Photoproduction has therefore been central to lepton–lepton and lepton–hadron colliders such as LEP and HERA.
The renewed interest in this process arises both from its relevance to the approved future Electron–Ion Collider (EIC) and from recent measurements in ultra-peripheral collisions (UPCs) at the LHC\,\cite{ATLAS:2024mvt}.
\\
To describe existing measurements and provide precise predictions for future experiments, it is essential to implement photoproduction of dijets in state-of-the-art Monte Carlo event generators such as \powhegbox\,\cite{Nason:2004rx,Frixione:2007vw,Alioli:2010xd}.
Matching next-to-leading order matrix elements with parton showers (NLO+PS) ensures both fixed-order accuracy for hard emissions and an adequate treatment of soft and collinear regions.
This significantly improves the reliability of predictions for exclusive observables.
Ultimately, such an implementation not only enhances the comparison with experimental data but also strengthens the extraction of photon and nuclear structure functions.
\\
In this contribution we summarize the key ingredients and present preliminary results for the implementation of dijet photoproduction in \powhegbox.

\section{Particle Spectra}
In the standard \powhegbox\ framework, initial hadron beams are resolved once into their partonic structure.
To accommodate the full structure required for photoproduction, namely extracting the photon flux from the lepton beam and, where relevant, the partonic substructure of the photon, we introduced modifications to the code.
Both levels of spectra are discussed in the following.

\subsection{Photon Spectrum in the Electron}
Fast-moving charges generate an electromagnetic field that, for small scattering angles, can be interpreted as the emission of a quasi-real (low-virtuality) photon.
This concept, originally introduced by Weizsäcker\,\cite{vonWeizsacker:1934nji} and Williams\,\cite{Williams:1934ad}, simplifies the calculation of the leptonic cross section drastically.
In the so-called equivalent photon or Weizsäcker-Williams approximation (WWA), it factorizes into a photon flux and a photonic cross section
\be
 \sigma_{e P}(e P \rightarrow e X)=\int_{0}^{1} \mathrm{d}x  F_{\gamma / e}(x) \sigma_{\gamma P}(\gamma P \rightarrow X)
\ee
with $x$ the energy fraction of the lepton transferred to the photon. 
In the \powhegbox\ implementation we used the improved WWA\,\cite{Frixione:1993yw} for lepton beams, including an additional mass term in the flux
\be
F_{\gamma / e}(x)=\frac{\alpha}{2 \pi}\left[\frac{1+(1-x)^{2}}{x} \ln \frac{Q_{\max }^{2}}{Q_{\min }^{2}}+2 m_{e}^{2} x\left(\frac{1}{Q_{\max }^{2}}-\frac{1}{Q_{\min }^{2}}\right)\right] \, .
\ee
Due to kinematical constraints, the photon virtuality is bounded from below by  $Q^2_{\text{min}}=m_e^2x^2/(1-x)$ while the upper limit can be set by the user either directly or indirectly through a maximum lepton scattering angle.

\subsection{Photon PDF}
\begin{figure}[h]
    \centering
    \includegraphics[width=\linewidth]{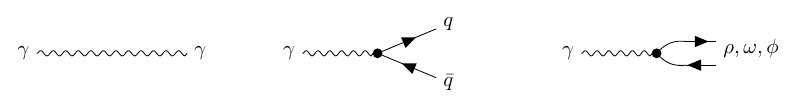}
    \caption{Sketch of the direct (left), point-like (middle) and hadronic structure (right) of the photon.}
    \label{fig:photonPDF}
\end{figure}

The structure of the photon consists of several components, as illustrated in \cref{fig:photonPDF}.
In a hard interaction, the photon may either participate as an elementary particle (direct contribution) or fluctuate into a hadronic state, thereby exposing its partonic content (resolved contribution).
The resolved part can be further divided into two subcategories: the point-like contribution, arising from the perturbative splitting of the photon into a quark–antiquark pair, and the non-perturbative component, described by the vector meson dominance (VMD) model.
In the latter, the photon is treated as a superposition of the pure photon and low-mass vector mesons with the same quantum numbers, such as the $\rho$, $\omega$, and $\phi$ mesons\,\cite{Nisius:1999cv}.
\\
Naturally, the photon PDFs fulfill a set of evolution equations which have the schematic form of
\be
\frac{d f^{i / \gamma}\left(x, Q^{2}\right)}{d \ln Q^{2}}=\frac{\alpha}{2 \pi} P_{i \leftarrow \gamma} \otimes f^{\gamma / \gamma}+\frac{\alpha_{s}}{2 \pi} \sum_{j} P_{i \leftarrow j} \otimes f^{j / \gamma} \, .
\ee
The second part of this equation resembles the familiar DGLAP evolution for proton PDFs, involving the strong coupling and the QCD splitting functions.
 In contrast, the first part is distinctive for photon PDFs and contains the scale-independent photon-in-photon contribution, represented by
 \be 
 f^{\gamma / \gamma}(x)=\delta(1-x) \, ,
 \ee
together with the electromagnetic splitting of the photon into QCD partons.
\\
Our implementation currently includes the GRV\,\cite{Gluck:1991ee}, as well as the CJKL PDF set\,\cite{Cornet:2002iy}.
When the original factorization scheme differs from $\overline{\text{MS}}$, the PDFs are transformed accordingly\,\cite{Gluck:1991ee}.

\section{Preliminary Results}
\begin{figure}[h]
     \centering
     \begin{subfigure}[b]{0.49\textwidth}
         \centering
         \includegraphics[width=\textwidth]{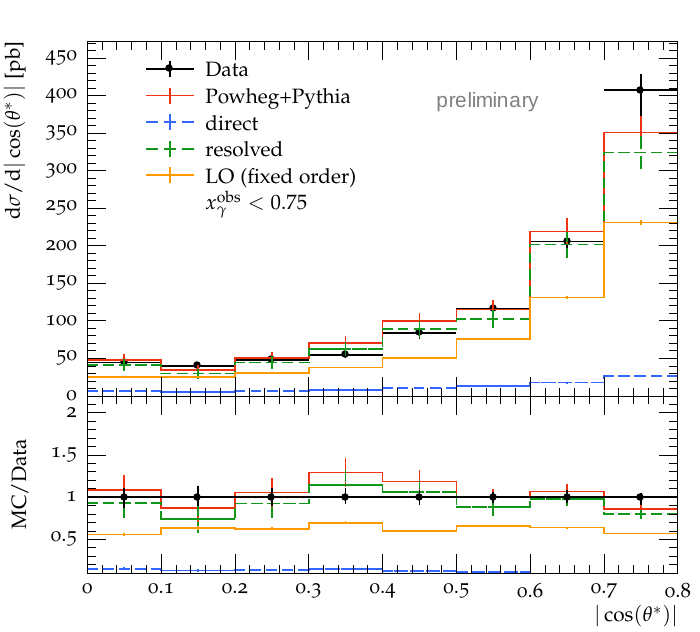}
         \caption{}
         \label{fig:}
     \end{subfigure}
     \begin{subfigure}[b]{0.49\textwidth}
         \centering
         \includegraphics[width=\textwidth]{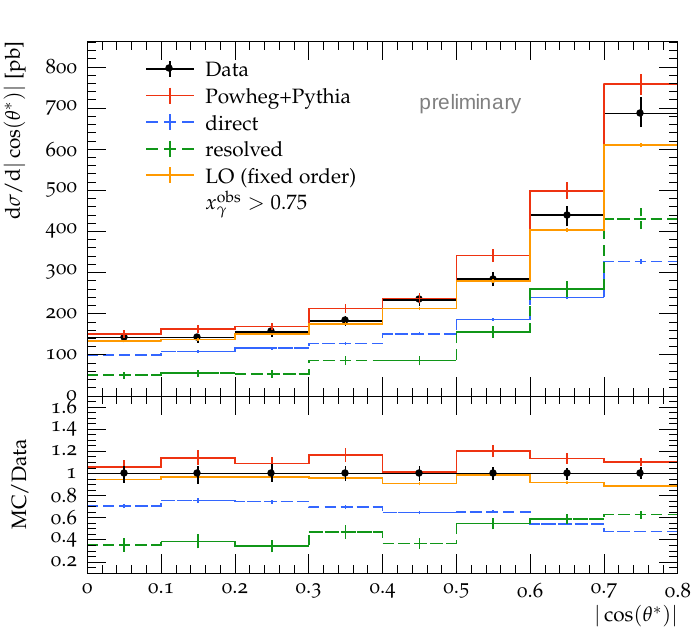}
         \caption{}
         \label{fig:}
     \end{subfigure}
    \caption{Preliminary NLO+PS results for photoproduced dijet events from \powhegbox\ compared to ZEUS data\,\protect\cite{ZEUS:2001zoq}. Differential cross sections are shown as a function of $\cos(\theta^*)$ for $x_\gamma^\text{obs} < 0.75$ (a) and $x_\gamma^\text{obs} > 0.75$ (b).}
        \label{fig:plot1}
\end{figure}

\begin{figure}[h]
     \centering
     \begin{subfigure}[b]{0.49\textwidth}
         \centering
         \includegraphics[width=\textwidth]{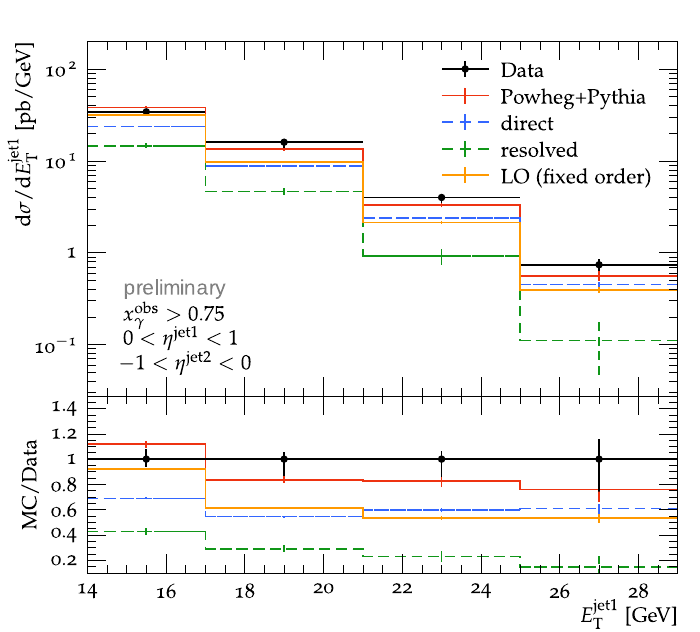}
         \caption{}
         \label{}
     \end{subfigure}
     \begin{subfigure}[b]{0.49\textwidth}
         \centering
         \includegraphics[width=\textwidth]{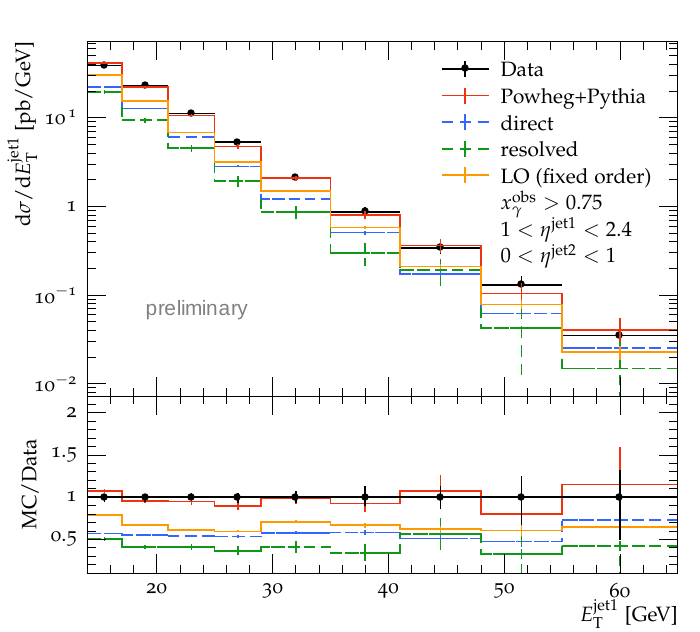}
         \caption{}
         \label{fig:}
     \end{subfigure}
        \caption{Preliminary NLO+PS results for photoproduced dijet events from \powhegbox\ compared to ZEUS data\,\protect\cite{ZEUS:2001zoq}. Differential cross sections are shown as a function of the transverse energy of the highest-$p_T$ jet for $x_\gamma^\text{obs} > 0.75$. Panel (a) corresponds to rapidity cuts $0 < \eta^{\text{jet1}} < 1$ and $-1 < \eta^{\text{jet1}} < 0$, while panel (b) corresponds to $1 < \eta^{\text{jet1}} < 2.4$ and $0 < \eta^{\text{jet1}} < 1$.}
        \label{fig:plot2}
\end{figure}

The matrix elements for the hard partonic subprocesses were implemented in \powhegbox\ using analytic expressions.
The direct contribution, obtained with \texttt{FeynArts}\,\cite{Hahn:2000kx} and \texttt{FeynCalc}\,\cite{Mertig:1990an,Shtabovenko:2016sxi,Shtabovenko:2020gxv}, was validated by cross-checking against \texttt{MadGraph}\,\cite{Alwall:2014hca}.
For the resolved contribution, we interfaced to the existing QCD dijet implementation\,\cite{Alioli:2010xa}, which had already been validated and was therefore not cross-checked again.
\\
Beyond these internal validations, we are carrying out a comparison with HERA data.
In the following, we present preliminary results for positron–proton collisions at beam energies of $E_e = \SI{27.5}{\giga\electronvolt}$ and $E_p = \SI{820}{\giga\electronvolt}$, compared with data recorded by the ZEUS detector\,\cite{ZEUS:2001zoq}.
The observables $x_\gamma^{\text{obs}}$ and $\cos\theta^*$ are defined as
\be
    x_{\gamma}^{\mathrm{obs}}=\frac{E_{T}^{\mathrm{jet} 1} e^{-\eta^{\mathrm{jet} 1}}+E_{T}^{\mathrm{jet} 2} e^{-\eta^{\mathrm{jet} 2}}}{2 x_{\gamma / e} E_{e}} 
    \quad \text{and} \quad
\cos \theta^{*}=\tanh \left(\frac{\eta^{\mathrm{jet} 1}-\eta^{\mathrm{jet} 2}}{2}\right) \, .
\ee
At fixed-order Born level, $x_\gamma^{\text{obs}}$ coincides with the momentum fraction of the parton in the photon.
In practice, it serves as an experimental proxy to separate direct and resolved contributions, with resolved events expected to show up at lower values while direct events appear close to one.
\\
In \cref{fig:plot1} we show results generated at leading order (LO) as well as at next-to-leading order (NLO) with additional final state radiation attached by \texttt{Pythia}\,\cite{Bierlich:2022pfr}.
As evident from the separated contributions, direct events play only a minor role when selecting $x_\gamma^{\text{obs}} < 0.75$.
The overall shape of the predictions follows the data reasonably well, although not all bins agree within statistical uncertainties.
In particular, for $x_\gamma^{\text{obs}} > 0.75$ the predictions tend to overshoot the measurements slightly.
\\
As a second example, we consider the transverse energy distribution of the highest-$p_T$ jet in \cref{fig:plot2}.
Already at LO, the predictions reproduce the overall shape of the measured cross section, indicating that the basic kinematics of the process are well captured.
Once NLO+PS corrections are included, the absolute normalization improves significantly and the predictions move closer to the measurement.
Especially in the right panel, we can report satisfactory agreement with the data.
In the shown regime, where $x_\gamma^{\text{obs}}>0.75$, the separated contributions reveal a clear dominance of the direct process, confirming yet again that $x_\gamma^{\text{obs}}$ is an effective discriminator between direct and resolved events.
\\
All results presented here are of preliminary nature, as the validation and consistency checks of the implementation are still ongoing.
They should therefore be interpreted with due caution until the code has been fully verified.
\section{Summary and Outlook}
A precise prediction of photoproduced dijet events at NLO+PS provides a powerful means to compare with and model experimental data.
Achieving this requires an accurate description of the photon flux in the lepton beam, suitable photon PDF sets, and the proper matching of NLO fixed-order calculations to parton showers.
These key elements have been briefly discussed in this work.
\\
Our framework may be used to create relevant predictions for upcoming EIC measurements.
Future work will also focus on linking and implementing accurate photon fluxes for hadron beams in order to compare to existing data from UPCs.

\section*{Acknowledgments}
This work has been supported by the BMBF under contract 05P24PMA.

\end{document}